\documentclass[a4paper,11pt]{article}
\usepackage{fullpage}
\usepackage{feynmf,amsmath}
\usepackage{epsfig}
\def\sign{\operatorname{sign}}
\def\tr{\operatorname{tr}}
\def\Tr{\operatorname{Tr}}
\def\diag{\operatorname{diag}}

\begin{document}

\begin{fmffile}{chrdgrms}

\title{{\bf Nonfactorizable contributions to the decay mode
 $D^0 \to K^0 \bar K^0$}}
\author{J. O. Eeg$^a$, S. Fajfer$^{b,c}$, and J. Zupan$^c$ \\
$^a$Dept. of Physics, Univ. of Oslo, N-0316 Oslo, Norway\\
$^b$Physics Department, University of Ljubljana, Jadranska 19,
SI-1000 Ljubljana, Slovenia\\
$^c$Institut Jo\v{z}ef Stefan, Jamova 39, SI-1000 Ljubljana, Slovenia}

\maketitle

\begin{abstract}
We point out that the decay mode $D^0 \rightarrow K^0 \bar K^0$ has no
factorizable contribution. In the chiral perturbation language, treating
$D^0$ as heavy, the ${\cal O}(p)$ contribution is zero. We calculate
the nonfactorizable  chiral
loop contributions of order ${\cal O}(p^3)$. Then, we use a heavy-light
type chiral quark model to calculate nonfactorizable  tree level terms,
 also of order ${\cal O}(p^3)$, proportional to the gluon condensate.
 We find that both the chiral  loops and the gluon condensate
contributions
 are of the same order of magnitude as the experimental amplitude.
\end{abstract}

PACS number(s): 13.25.-k, 13.25.Ft, 13.39.Fe, 12.39.Hg

\section{Introduction}
The decay mechanism of the weak nonleptonic $D^0$ decays has motivated
numerous studies   \cite{WSB}-
%XYP,DDL,HL,GPW,KUSV,Z,T,BGR,
\cite{BLMPS}.
For nonleptonic decays of  $D$ mesons, as well as  for $K$'s and $B$'s,
the so called {\em factorization} hypothesis has been commonly used.
For nonleptonic decays, the effective Lagrangian at quark level
has the form
\begin{equation}
 {\cal L}_{W} \; = \; \sum_i \, C_i \, Q_i \;
 \label{LCQ},
\end{equation}
where the coefficients $C_i$ contain all the short distance
electroweak and QCD effects to a certain order in perturbation theory,
and the $Q_i$'s are quark operators. Typically, these quark operators
are
products of (pseudo) scalar- or vector- currents: $Q = j(1) \, j(2)$.
Then, for a nonleptonic decay $M \rightarrow M_1 + M_2$,  the
factorization
hypothesis (-also called vacuum saturation approximation)
 gives prescriptions of the form
\begin{equation}
\langle  M_1 \, M_2| Q |M\rangle  \; \; \rightarrow \; \;
\langle  M_1| j(1) |0\rangle  \langle M_2 | j(2) | M\rangle  \; .
\label{fact}
\end{equation}
The factorization hypothesis are known to fail badly for nonleptonic 
$K$ decays \cite{K-fact,BEF,cheng}.
On the
other hand, there are  certain heavy hadron weak decays where
factorization might apply.
Recently,  the understanding of factorization  for
exclusive nonleptonic decays of $B$ mesons in terms of QCD in the
 heavy quark limit has
been considerably improved  \cite{BBNS}.
In this paper we will discuss nonfactorizable terms for $D$ decays,
in particular for the decay mode $D^0\to K^0\bar{K^0}$.

 Even though the factorization hypothesis might work
reasonably well if one is interested in an order of magnitude estimate,
 it does not reproduce experimental data completely. For example, a
naive
application of  factorization  in charm decays
leads to  rates for  the $D^0 \to \pi^0
\bar K^0$,
$D^0 \to \pi^0 \pi^0$, $D^0 \to K^+ K^-$  $D^0 \to \pi^+ \pi^-$
decays which are
too strongly suppressed.
Moreover, and this is the important point of this paper:
 in $D^0\to K^0\bar{K^0}$,  factorization   misses
completely,
predicting a vanishing branching ratio, in contrast with the
experimental
 situation.

To see this, note that at tree level the $D^0\to K^0\bar{K^0}$ decay
might
occur due to two annihilation diagrams \cite{WSB} which could
potentially
create the $K^0 \bar K^0$ state. However, they cancel each other by the
GIM
 mechanism. Moreover, in factorization limit, the amplitude is
proportional to
\begin{equation}
\langle  K^0 \bar K^0| V_\mu |0\rangle  \langle 0 | A^\mu | D^0\rangle 
\simeq (p_{K^0} - p_{\bar K^0})_\mu \,  f_D p_{D}^\mu= 0 \; .
\label{eq-9}
\end{equation}
In many of the studies (e.g.
\cite{XYP,DDL,HL,GPW,Z}) this decay has been understood as a
result of final state interactions (FSI).
In the analysis of ref. \cite{XYP}
the rescattering mechanism included $K^+ K^-$ and $\pi^+ \pi^-$
states leading to a branching ratio
$B(D^0 \to K^0 \bar K^0)$ $= \frac{1}{2} B(D^0 \to K^+ K^-)$.
Experimental data on the other hand are  \cite{PDG-00} $B(D^0 \to K^0
\bar K^0)
= (6.5\pm 1.8)\times 10^{-4}$ and $B(D^0 \to K^+ K^-)=
 (4.25\pm 0.16)\times 10^{-3}$.
 A recent investigation of the $D^0\to K^0\bar {K^0}$ decay mode
performed in
\cite{DDL} has focused on the  $s$ channel  and the $t$ channel one
particle exchange contributions. The $s$ channel contribution has been
taken into account through the poorly  known scalar meson $f_0(1710)$
and was found to be very small, while the one particle $t$-exchanges
yielded higher contributions, with pion exchange being the highest. In
the approach of \cite{T} the  $D^0 \to K^0 \bar K^0$ was
realized through the scalar glueball
or glue-rich scalar meson.

The second instructive case concerning the factorization hypothesis,
is offered by the analyses of nonleptonic $K$ meson decays. Namely,  it
is well known that  factorization does not work in  nonleptonic $K$
decays. Among many approaches the Chiral Quark Model
 ($\chi$QM) \cite{pider} was shown to be able to accommodate the 
intriguing  $\Delta I = 1/2$ rule  in
$K \to \pi \pi$ decays,  as well
as CP violating parameters, by systematic involvement of the soft
gluon emission forming gluon condensates and chiral
loops at ${\cal O} (p^4)$ order \cite{BEF}.
In the $\chi$QM \cite{chiqm},
the light quarks ($u, d, s$) couple to the would-be Goldstone octet
mesons ($K, \pi, \eta$) in a chiral invariant way, such that all effects
are in principle calculable in terms of physical quantities and a few
model dependent parameters, namely the quark condensate, the gluon
condensate
and the constituent quark mass \cite{BEF,pider,epb}.
Also in ``generalized factorization'', it was shown  \cite{cheng} that
 the inclusion of gluon condensates  is important
 in order to understand  the $\Delta I = 1/2$ rule
for $K \rightarrow 2 \pi$ decays.

As the $\chi$QM  approach successfully indicated the main
 mechanisms in $K \to \pi \pi$
decays, it seems worthwhile to investigate decays of charm mesons
within a similar framework. However, in the case of $D$ meson decays one
has to extend the ideas of the $\chi$QM to the sector involving a
heavy quark ($c$) using the chiral symmetry of light degrees of freedom as
well as heavy quark symmetry and
Heavy Quark Effective Field Theory (HQEFT).
  Such ideas have already been presented
in previous papers \cite{barhi,effr,itCQM} and lead to the formulation of
  Heavy-Light Chiral Quark Models (HL$\chi$QM).
In our formulation of the  HL$\chi$QM Lagrangian, an unknown coupling
constant
appears in  the term  that couples  the heavy meson to a heavy and
 a light quark.
Our strategy is  to relate  expressions involving this coupling to
physical
 quantities, as it is done within the $\chi$QM \cite{BEF}. We perform
the bosonization by integrating out the light and heavy quarks and 
obtain
a heavy quark symmetric chiral Lagrangian involving light and heavy
mesons \cite{itchpt,wise}.

Because the ${\cal O}(p)$ (factorizable) contribution is zero
as seen in Eq. \eqref{eq-9},
we will  try in this paper to approach to the $D^0 \to K^0 \bar K^0$
decay
systematically to ${\cal O}(p^3)$. We do this by including first
 the nonfactorizable
 contributions coming from the chiral loops. These are based on the weak
 Lagrangian
corresponding  to the factorizable  ${\cal O}(p)$ terms for
$D^0 \rightarrow \pi^+ \pi^-$ and $D^0 \rightarrow K^+ K^-$.
(Note that the velocity $v^\mu = p^\mu_D/m_D$ is considered to be
${\cal O}(p^0)$ in the chiral counting).
Second, we consider the  gluon condensate contributions, also of
${\cal O}(p^3)$ within the $\chi$QM and HL$\chi$QM framework.
It should be noted that because the energy release in
$D \rightarrow K \bar{K}$ is of order one GeV (-in contrast to 200 MeV for
$K \rightarrow 2 \pi$), the next to leading ${\cal O}(p^5)$ terms might
be almost
of the same size numerically compared to our ${\cal O}(p^3)$ terms.
Still, the amplitude of $D^0 \to K^0 \bar K^0$ calculated
within the framework of ${\cal O}(p^3)$, has a reliable order of
magnitude.
Note that we have also omitted $1/m_Q$ terms in the framework of HQEFT.

Our paper is organized as follows:
In Section 2 we write down the basic Lagrangians including the weak
Lagrangian at quark level (with special emphasis on the terms giving
rise to
the nonfactorizable gluon condensate contributions) as well as
 the standard strong chiral Lagrangians
for the light and heavy meson sectors.
The chiral loop contributions to the decay amplitudes are presented in
Section 3.
 The details of the
Heavy - Light Chiral Quark  Model (HL$\chi$QM) are presented in
Section 4, while the bosonization of the weak quark currents is given in
Section 5. The results are given in section 6.
Appendix A  contains some details from the chiral loop integrals,
Appendix B some details about  the $D$ meson decay constant, while
 Appendix C contains some loop integrals within the HL$\chi$QM.

\section{Basic Lagrangians}

The effective weak Lagrangian at quark level
relevant for  $D \rightarrow \pi \pi, K \bar{K}$ is
\begin{equation}
 {\cal L}_{W}= \widetilde{G} \left[ c_A \, (Q_A -Q_C) \;
+ c_B \, (Q_B^{(s)} - Q_B^{(d)}) \right] \; ,
 \label{Lquark}
\end{equation}
where
$\widetilde{G} =  -   \, 2 \sqrt{2} G_F V_{us}\,V^*_{cs}$, and
\begin{eqnarray}
Q_A  =   ( \overline{s}_L \gamma^\mu  c_L )  \; \,
           ( \overline{u}_L \gamma_\mu  s_L )
\; \;  , \,
Q_{C}  =  ( \overline{d}_L \gamma^\mu c_L ) \; \,
          ( \overline{u}_L  \gamma_\mu d_L )
\, , \nonumber  \\
Q_B^{(q)}  =   \, ( \overline{u}_L \gamma^\mu c_L ) \; \,
           ( \overline{q}_L \gamma_\mu q_L )
\; \; , \qquad
 (q  \;  = \; s,d)
\, ,
\label{QA-QC}
\end{eqnarray}
are quark operators.

Using Fierz transformations and the
following relation between the generators of $SU(3)_c$ ($i,j,l,n$
are color indices running from 1 to 3 and $a$ is an index
 running over the eight gluon charges):
\begin{equation}
\delta_{i j}\delta_{l n} \, = \, \frac{1}{N_c} \delta_{i n} \delta_{l j}
 \; + \; 2 \; t_{i n}^a \; t_{l j}^a \; ,
\label{fierz}
\end{equation}
one obtains
\begin{equation}
\begin{split}
Q_A  \; = \; \frac{1}{N_c} \,  Q_B^{(s)} \, + \, R_B^{(s)}
\; , &\;
Q_C  \; = \; \frac{1}{N_c} \,  Q_B^{(d)} \, + \, R_B^{(d)} \; ,\\
Q_B^{(s)}  \; = \; \frac{1}{N_c} \,  Q_A \, + \, R_A
\; ,&\; 
Q_B^{(d)}  \; = \; \frac{1}{N_c} \,  Q_C \, + \, R_C \; ,
\label{QFierz}
\end{split}
\end{equation}
where the $R$'s  correspond to color
exchange between two currents and
is genuinely nonfactorizable:
\begin{eqnarray}
R_A \, = \, 2 \, ( \overline{s}_L \, \gamma^\mu \, t^a \, c_L \, ) \;
( \overline{u}_L \, t^a \, \gamma_\mu \, s_L \,) \; , \qquad
R_C \, = \, 2 \, ( \overline{d}_L \, \gamma^\mu \, t^a \, c_L \, ) \;
( \overline{u}_L \, t^a \, \gamma_\mu \, d_L \,) \; ,
\nonumber \\
R_B^{(q)} \, = \, 2 \, ( \overline{u}_L \, \gamma^\mu \, t^a \, c_L \, )
\;
( \overline{q}_L \, t^a \, \gamma_\mu \, q_L \,)
 \; \, ;  \qquad
 (q  \;  = \; s,d)
\, ,
\label{FiCo}
\end{eqnarray}

The operators can be written in terms of currents, for instance:
\begin{eqnarray}
Q_B^{(s)} - Q_B^{(d)} = \, J^Y_\mu \; j^\mu_X \; \; \; , \; \;
R_B^{(s)} - R_B^{(d)} = \,2 \,  J^{Y,a}_\mu \; j^{\mu,a}_X  \; \,  ,
\label{QBsd}
\end{eqnarray}
where
\begin{eqnarray}
J^{Y,a}_\mu \, \equiv \,   \overline{u}_L \, \gamma^\mu \, t^a \, c_L  
\; ;
\qquad
j^{\mu,a}_X \, \equiv  \,\overline{s}_L \, t^a \, \gamma^\mu \, s_L \,
- [s \rightarrow d] \;  \; \, .
\label{eq-10}
\end{eqnarray}
The currents without color index are given by the corresponding
expressions
dropping the color matrix.

The factorization approach  amounts to writing the currents
$J_\mu^Y$, $j_\mu^X$ in terms of hadron (in our case meson)
fields only, so that the operator $Q_B^{(s)} - Q_B^{(d)}$  in the
left equation of \eqref{QBsd} is equal to the product of two
meson currents. The color currents in \eqref{QBsd} are then zero if
hadronized (mesons are color singlet objects).
There is also a replacement of the Wilson coefficients in the hadronized
effective
weak Lagrangian $c_{A,B}\to c_{A,B}(1+1/N_c)$. 
Combining heavy quark symmetry and chiral symmetry of the light sector,
we can obtain the weak chiral Lagrangian for nonleptonic 
$D$ meson decays due to 
factorizable terms. Then we can first use this to 
calculate  nonfactorizable contributions due to chiral loops.
 Second, we can  calculate
 the color currents' contribution  using the
gluon  condensate within
the framework of the HL$\chi$QM.

Treating the light pseudoscalar mesons as pseudo-Goldstone bosons one
obtains
the  usual ${\cal O}(p^2)$ chiral Lagrangian
\begin{equation}
{\cal L}_{\text str}^{(2)}=\frac{f^2}{8} \tr (\partial^\mu \Sigma
\partial_\mu \Sigma^\dagger)+\frac{f^2 B_0}{4}\tr({\cal M}_q \Sigma
+{\cal M}_q\Sigma^\dagger) \; ,\label{eq-11}
\end{equation}
where $\Sigma = \exp{(2 i \Phi/f)}$ with
$\Phi =  \sum_j \lambda^j \pi^j$ containing the Goldstone bosons $\pi,
K,
\eta$, while the trace $\tr$ runs over flavor indices and ${\cal
M}_q=\diag
(m_u,m_d,m_s)$ is the current quark mass matrix. From this Lagrangian,
 we can deduce the
 light weak current  to  ${\cal O}(p)$
\begin{equation}
j_\mu^X \, = \, -i\frac{f^2}{4}\tr(\Sigma \partial_\mu
\Sigma^\dagger\lambda^X) \; ,
\label{jX}
\end{equation}
corresponding to the quark current
$j_\mu^X=\bar{q}_{L}\gamma_\mu\lambda^Xq_{L}$. ($\lambda^X$ is a SU(3)
flavor matrix.)

In the heavy meson sector interacting with light mesons we have
the following lowest order ${\cal O}(p)$  chiral Lagrangian
\begin{equation}
{\cal L}_{\text str}^{(1)}=-\Tr(\bar{H}_{va}iv\cdot D_{ab}H_{vb})-g
\Tr(\bar{H}_{va}H_{vb} \gamma_\mu {\cal A}_{ba}^\mu \, \gamma_5) \; ,
\label{eq-8}
\end{equation}
where $i D_{ab}^\mu H_b=i \partial^\mu H_a - H_b{\cal V}_{ba}^\mu$,
 the trace $\Tr$ runs over Dirac indices.
Note that in (\ref{eq-8}) and the rest of this section
$a$ and $b$ are {\em flavor} indices.

The vector and axial vector fields
${\cal V}_{\mu}$ and
${\cal A}_\mu$ in (\ref{eq-8}) are given by:
\begin{equation}
{\cal V}_{\mu} \equiv \frac{i}{2}(\xi\partial_\mu\xi^\dagger
+\xi^\dagger\partial_\mu\xi) \qquad ;  \qquad
{\cal A}_\mu \equiv \frac{i}{2}
(\xi\partial_\mu\xi^\dagger -\xi^\dagger\partial_\mu\xi) \; ,
\label{defVA}
\end{equation}
where $\xi = \exp{(i \Phi/f)}$. The heavy meson field
 $H_{va}$ contains
 a spin zero and spin one boson:
\begin{equation}
H_{va} \equiv P_+ (P_{\mu a} \gamma^\mu -
i P_{5 a} \gamma_5)\; ,
\end{equation}
\begin{equation}
\overline{H}_{va}
=\gamma^0 (H_{va})^\dagger \gamma^0
=\left[ P_{\mu a}^{\dagger} \gamma^\mu
- i P_{5 a}^\dagger \gamma_5\right] P_+ \; ,
\label{barH}
\end{equation}
with $ P_{\pm}=(1 \pm \gamma^\mu v_\mu)/2 $ being the the  projection operators.
The field $P_5(P_5^\dagger)$ annihilates (creates) a pseudoscalar 
meson  with
a heavy quark  having  velocity $v$,
and similar for spin one mesons.

For a decaying heavy quark, the
 weak current is given by
\begin{equation}
J_a^\lambda =\overline{q}_a \gamma^\lambda L Q \; ,
\label{Lcur}
\end{equation}
where $L=(1-\gamma_5)/2$ and $Q$ is the heavy quark field in the
 full theory, in our case a $c$-quark, and $q$ is the light quark field.
(For flavor $a=u$, this is the current $J_\mu^Y$ in (\ref{QBsd}).)

From symmetry grounds,
 the heavy-light weak current is to  ${\cal O}(p^0)$ bosonized
in the following way \cite{wise}
\begin{equation}
J^\lambda_a =  \alpha_H \Tr [\gamma^\lambda \, L \, H_{vb}\,
\xi_{ba}^\dagger]
\; ,
\label{JH}
\end{equation}
where $\alpha_H$ is related to the
 physical decay constant $f_D$ through the well known matrix element
\begin{equation}
\langle  0| \overline{u}\gamma^\lambda \gamma_5 c|D^0 \rangle 
= -2 \langle  0| J_a^\lambda|D^0 \rangle 
=im_D v^\lambda f_D \; .
\label{fD}
\end{equation}
Note that the current (\ref{JH}) is  ${\cal O}(p^0)$ in the chiral
counting.

\section{Chiral loop contributions}

In the factorization limit there are no contributions
to $D^0\to K^0 \bar{K^0}$ at tree level. The observation of a partial
decay width  $B(D\to K^0\bar{K^0})=(6.5\pm 1.8)\times 10^{-4}$ on the
other hand implies that we can expect sizable contributions at the one
loop level. Calculations to one loop  in the framework of combined chiral
perturbation theory and HQEFT involves  a
construction of the most general effective Lagrangian that has the
correct
symmetry properties in order to make the renormalization work.
 We  discuss  constructions of counterterms in the end of Sect. 5.

We work in the strict $\overline{MS}$ renormalization scheme, where we
put
$\bar{\Delta}=\frac{2}{\epsilon}-\gamma+\ln(4\pi)+1$ equal to
one  in the loop calculations. This choice, $\bar{\Delta}=1$, determines
the
appropriate renormalization of couplings in the ${\cal O}(p^3)$ effective
Lagrangian and is the same as made by Stewart in \cite{stewart}, while
it differs from the one used by authors of Ref. \cite{itchpt}, who
use $\bar{\Delta}=0$.  We consider
 only contributions coming from  the $c_A$ part of the weak Lagrangian
 as $c_B$ is suppressed compared to $c_A$ \cite{buras}.

\begin{figure}
\begin{center}
\epsfig{file=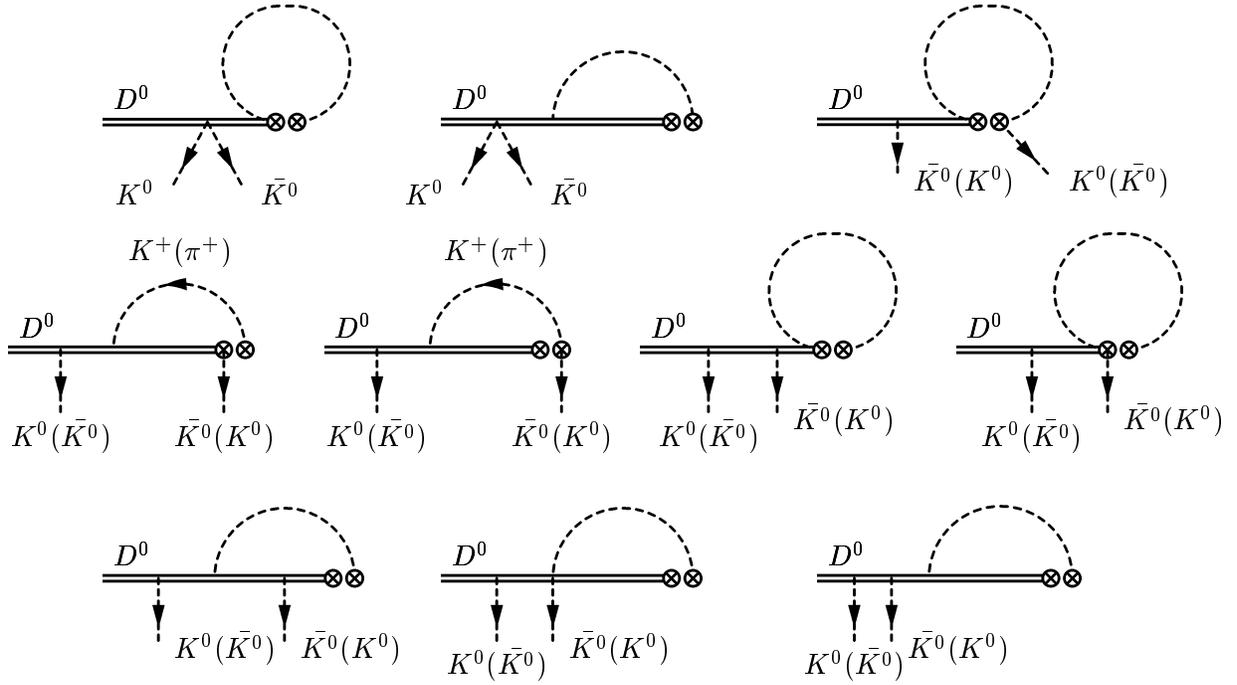}
\caption{Diagrams that give zero contribution since  the relevant
vertices
appearing in the heavy meson chiral Lagrangian \eqref{eq-8} are zero.
 The double line represents
heavy meson $D$ or $D^*$, while dashed lines denote pseudo-Goldstone
bosons. }\label{fig-2}
\end{center}
\end{figure}

\begin{figure}
\begin{center}
\epsfig{file=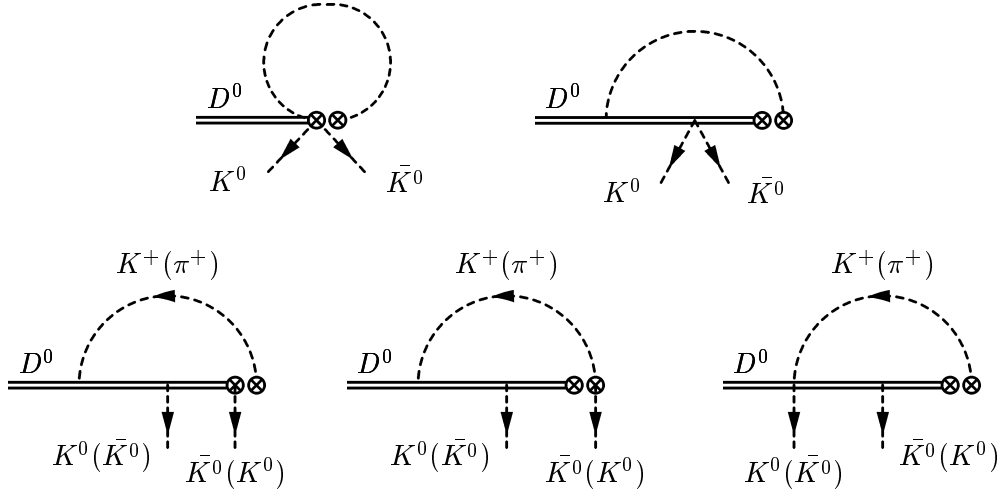}
\caption{Diagrams that give zero contributions since the loop
integrals are zero.The double line represents
heavy meson $D$ or $D^*$, while dashed lines denote pseudo-Goldstone
bosons.}\label{fig-3}
\end{center}
\end{figure}

\begin{figure}
\begin{center}
\epsfig{file=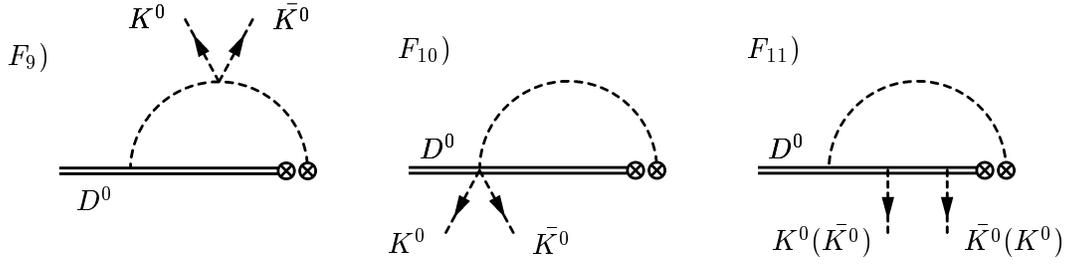}
\caption{Power suppressed diagrams (neglected in  the
calculation).}\label{fig-4}
\end{center}
\end{figure}

\begin{figure}
\begin{center}
\epsfig{file=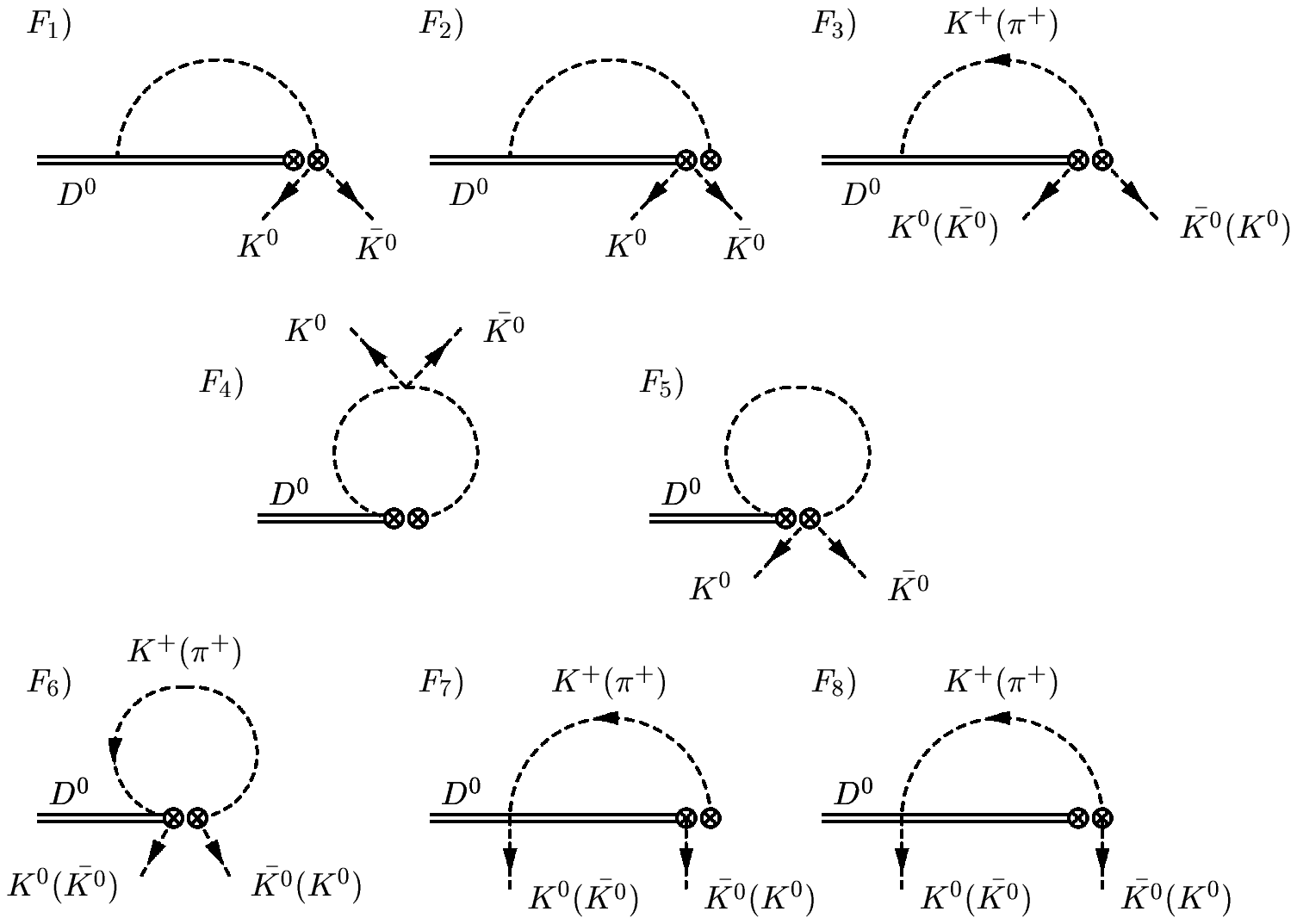}
\caption{The nonzero diagrams in $D^0 \to K^0 \bar{K^0}$
decay.}\label{fig-5}
\end{center}
\end{figure}

Writing down the most general one loop graphs with two outgoing
Goldstone
bosons ($K^0$ and $\bar{K^0}$)  one arrives at 26 Feynman
diagrams. A number of these give zero contributions and are shown on
Figures \ref{fig-2},\ref{fig-3},\ref{fig-4}, while the graphs that do
contribute to $D^0\to K^0\bar{K^0}$ decay are shown on Fig. \ref{fig-5}.
Note that factorizable loops which renormalize  vertices are omitted
(they do appear, however, in the loop determination of the $\alpha_H$
coupling
related to $f_D$. See Appendix B.)

To shorten the notation, the common factors in the $S$ matrix have been
organized such that the amplitude is written
\begin{equation}
{\cal M}(D^0 \rightarrow K^0 \bar{K^0}) \; = \; - \,
\frac{G_F}{\sqrt{2}} \, c_A  \, V_{us}\,V^*_{cs} \,
\frac{F}{8\pi^2}\sqrt{m_D}\; ,\label{eq-12}
\end{equation}
where
$F=\sum_n F_n$ is the sum of the amplitudes corresponding to the graphs
on Figs.
\ref{fig-2}, \ref{fig-3}, \ref{fig-4}, \ref{fig-5}. The partial decay
width
for the decay $D^0 \to K^0\bar{K^0}$ is then
\begin{equation}
\Gamma_{D^0\to
K^0\bar{K^0}}=\frac{1}{2\pi}\frac{G_F^2}{8m_D}
c_A^2  |V_{us} \, V^*_{cs}|^2
\frac{|F|^2}{(8\pi^2)^2}
p \; ,\label{eq-2}
\end{equation}
where $p$ is the $K^0(\bar{K^0})$ three-momentum in the $D^0$ rest frame
\begin{equation}
p=\frac{1}{2}\sqrt{m_D^2-4m_K^2} \; \, .
\end{equation}
The nonzero amplitudes corresponding to the graphs on Fig. \ref{fig-5}
are
\begin{align}
F_1&+F_2+F_3=-\frac{g \alpha_H}{f^2}\frac{13}{4} [ \Delta_d^* J_1(m_\pi
,\Delta_d^*)-\Delta_s^* J_1 (m_K,\Delta_s^*)],\label{eq-3}\\
\begin{split}
F_4&=-\frac{\alpha_H}{3 f^2}\frac{m_D}{2} \Bigl\{(m_D^2-2
m_K^2)[N_0(m_\pi,m_D^2)-N_0(m_K,m_D^2)]\Bigr.+\\
&\qquad\qquad+m_D^2[N_2(m_\pi,m_D^2)-N_2(m_K,m_D^2)]+\\
&\qquad\qquad\Bigl.+[N_3(m_\pi,m_D^2)-N_3(m_K,m_D^2)]-(m_\pi^2
-m_K^2)N_0(m_\pi,m_D^2)\Bigr\},\label{eq-5}
\end{split}\\
F_5&+F_6=\frac{\alpha_H m_D}{f^2}\frac{7}{24}[I_1(m_\pi)-I_1(m_K)],\\
\begin{split}
F_7&+F_8=\frac{\alpha_H}{4
f^2}\Bigl\{\Bigr.\tilde{\Delta}_d\bigl(J_1(m_K,\tilde{\Delta}_d)+J_2(m_K,\tilde{\Delta}_d)\bigr)-\tilde{\Delta}_s\Bigl(J_1(m_\pi,\tilde{\Delta}_s)+J_2(m_\pi,\tilde{\Delta}_s)\bigr)\\
&\qquad\qquad+m_D\frac{\Delta_d}{\tilde{\Delta}_d}I_2(m_K,\tilde{\Delta}_d)-m_D\frac{\Delta_s}{\tilde{\Delta}_s}I_2(m_\pi,\tilde{\Delta}_s)+\frac{m_D}{2\tilde{\Delta}_d}I_1(m_K)-\frac{m_D}{2
\tilde{\Delta}_s}I_1(m_\pi)\Bigl.\Bigr\},
\end{split}\label{eq-4}
\end{align}
where $\Delta_q^{(*)}=m_{D_q^{(*)}}-m_{D^0}$ and
$\tilde{\Delta}_q=m_D/2+\Delta_q$ for $q=d,s$.
Note that $\tilde{\Delta}_q$ are
of the order  $m_D/2$, a consequence of relatively high momenta
flowing in the loops of graphs $F_7, F_8$. The functions $I_1(m)$,
$I_2(m,\Delta)$, $J_1(m,\Delta)$, $J_2(m,\Delta)$, $N_0(m,k^2)$,
$N_2(m,k^2)$, $N_3(m,k^2)$ appearing in the  amplitudes
(24-27) can be found in Appendix \ref{app-A}.

It should be noted that in eq. (23-26)  all the
expressions  vanish in the exact $SU(3)$ limit, where $m_K \to
m_{\pi}$ and $\Delta_s \to \Delta_d$, $\tilde{\Delta}_s \to
\tilde{\Delta}_d$. This shows explicitly that the $D^0 \to K^0
\bar{K^0}$
decay mode is a manifestation of $SU(3)$ breaking effects (as already
noted by H. Lipkin \cite{HL}, if $U$ symmetry is exact, then
$\Gamma(D^0\to K^0\bar K^0)=0$).

The amplitudes  shown on Figs.~\ref{fig-2},\ref{fig-3},\ref{fig-4} are
either exactly zero or are suppressed by powers of $1/m_D$ and
$g=0.27$. The amplitudes corresponding to diagrams on
 Figs.~\ref{fig-2},\ref{fig-3} are zero due to symmetry reasons 
(because there are no such
couplings in the heavy sector chiral Lagrangian \eqref{eq-8}, or because
of Lorentz covariance), while the amplitudes $F_9$, $F_{10}$ and
$F_{11}$ shown on Fig.~\ref{fig-4} are power suppressed.
An analysis of the loop integrals leads to the conclusion
 that $F_9\sim g \bigl(\tilde{q}/m_D\bigr)^2 F_4$, $F_{10}\sim g
\bigl(\tilde{q}/m_D\bigr) F_4$ and $F_{11}\sim g^3
\bigl(\tilde{q}/m_D\bigr) F_4$,
where $\tilde{q}$ is a typical loop momentum  less that $m_D/2$,- so the
suppression need not be substantial. However,
a direct evaluation of the amplitude
$F_{10}$
shows that it is about  20 times smaller than $F_4$.
Therefore, in our numerical calculation we  neglect contributions
of $F_9$, $F_{10}$ and $F_{11}$.

\begin{figure}
\begin{center}
\epsfig{file=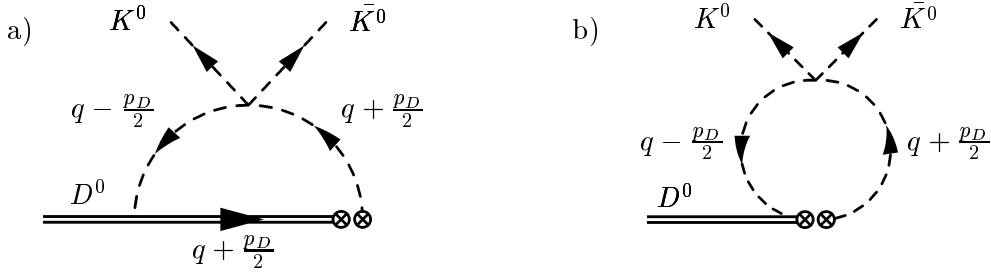}
\caption{The momenta flowing in the graphs corresponding to a) power
suppressed $F_9$ amplitude
 and b) the leading contribution $F_4$
amplitude. }\label{fig-1}
\end{center}
\end{figure}

\section{A Heavy-Light Chiral Quark Model
(HL$\chi$QM)}\label{HeavyLight}

The nonfactorizable contributions to $D^0\to K^0\bar{K^0}$ coming from
the
 chiral loop correction at the meson level obtained  in the previous
section
 are not the only contributions to ${\cal O}(p^3)$.
 In the effective weak Lagrangian
 \eqref{Lquark} there are, after Fierz transformations, terms that
involve
 color currents (see \eqref{QBsd},\eqref{eq-10}). As mesons are color
 singlet objects, the product of color currents does not contribute at
meson level in the factorization limit. However, at quark level they do
 contribute through  the gluon
 condensate as will be shown in  the next section. In order to
 estimate this contribution we have to establish the connection between
the underlying quark-gluon dynamics and the meson level picture. This is
 done through the use of the Heavy-Light Chiral Quark Model
(HL$\chi$QM).

Our starting point is the following Lagrangian containing both quark
 and meson fields:
\begin{equation}
{\cal L}={\cal L}_{HQ}+{\cal L}_{\chi QM} + {\cal L}_{Int} \; ,
\label{totlag}
\end{equation}
where
\begin{equation}
{\cal L}_{HQ}=\overline{Q}_v \, i v \cdot D \, Q_v
+ {\cal O}(m_Q^{-1})
\label{LHQET}
\end{equation}
is the Lagrangian for Heavy Quark Effective Field Theory (HQEFT).
The heavy quark field  $Q_v$
annihilates  a heavy quark  with velocity $v$ and mass $m_Q$. 
$D_\mu$ is the covariant derivative containing the gluon field.
The light quark sector is described by the Chiral Quark Model
($\chi$QM):
\begin{equation}
{\cal L}_{\chi QM}= \bar{q}(i\gamma^\mu D_\mu -{\cal M}_q) q
\, - \,  m_\chi \, (\bar{q}_R \Sigma^{\dagger} q_L \, +
 \, \bar{q}_L \Sigma q_R) \; ,
\label{chqmU}
\end{equation}
where $q =(u,d,s)$ are the light quark fields. The left and right-handed
 projections $q_L$ and $q_R$ are transforming under $SU(3)_L$ and
$SU(3)_R$
respectively. ${\cal M}_q$ is the current quark mass matrix, and
$\Sigma$
 is a 3 by 3 matrix containing
 the (would be)  Goldstone octet ($\pi, K, \eta$), appearing already in
\eqref{eq-11}.
The quantity $m_\chi$ is interpreted as the (SU(3)-invariant)
constituent
 quark mass for light quarks, supposed to appear due to chiral symmetry
 breakdown at a scale $\Lambda_\chi \, \sim $ 1 GeV.

 The $\chi$QM has a ``rotated version''
with  flavor rotated quark fields $\chi$ given by:
\begin{equation}
\chi_L = \xi^{\dagger} q_L \quad ; \qquad \chi_R = \xi q_R \quad ;
\qquad
\xi \cdot \xi = \Sigma \; .
\label{rot}
\end{equation}
In the rotated version, the chiral interactions are rotated  into the
kinetic term while the interaction term (proportional to $m_\chi$ in
(\ref{chqmU}) and responsible for the $\pi$ -quark couplings)
 become a pure (constituent) mass term:
\begin{equation}
{\cal L}_{\chi QM}=
\bar{\chi} \left[\gamma^\mu (i D_\mu \, + \, {\cal V}_{\mu} \,+\,
\gamma_5  {\cal A}_{\mu}) \,  - \, m_\chi \right]\chi \,
 -  \, \bar{\chi} \widetilde{M_q} \chi \; \, ,
\label{chqmR}
\end{equation}
and  $\widetilde{M_q}$ defines  the rotated version of the current
 mass term:
\begin{equation}
\widetilde{M_q} \equiv  \xi^{\dagger} {\cal M}_q \xi^{\dagger} \, R
\, + \, \xi {\cal M}_q^{\dagger} \xi \, L \;
\equiv  \; \widetilde{M}_q^R \, R \, + \, \widetilde{M}_q^L \, L
\;
\equiv  \; \widetilde{M}_q^V \, + \, \widetilde{M}_q^A \gamma_5  \; ,
\label{cmass}
\end{equation}
where  $L=(1-\gamma_5)/2$ is the left-handed projector in Dirac space,
and $R$ is the corresponding right-handed projector.
The Lagrangian (\ref{chqmR}) is manifestly invariant under the  unbroken
symmetry $SU(3)_V$ (if ${\cal M}_q$ is formally chosen to transform as
$\Sigma$).
In the light sector, the various pieces of the strong chiral Lagrangian
\eqref{eq-11}
can be obtained by integrating out the constituent quark fields $\chi$.
 This is
the  {\em bosonization} to be discussed in more detail in
the next section.

Similarly, a left handed current can be written ($\lambda^X$ is a SU(3)
flavor matrix)
\begin{equation}
\bar q_L \gamma^\mu \lambda^X q_L \, = \,  \,
\bar \chi_L  \gamma ^\mu \Lambda^X  \, \chi_L  \; ;
\qquad
\Lambda^X \, \equiv \, \xi^\dagger \lambda^X \,  \xi \; .
\label{leftcur}
\end{equation}
By coupling the fields ${\cal{A}}_{\mu},\, \widetilde M_q^{V,A}, \,
\Lambda^X$ to quark loops, the chiral Lagrangians of the weak sector
 can be obtained.

In the heavy-light case, the generalization of the
 meson-quark interactions in the pure light sector  $\chi$QM
is given by the following $SU(3)_V$
invariant Lagrangian \cite{barhi,effr,itCQM,AJ}:
\begin{equation}
{\cal L}_{Int} =
- G_H \, \left[ \bar{\chi}_f \, \overline{H}_{vf} \, Q_v \,
 +  \, \overline{Q}_v \, H_{vf} \, \chi_f \right] \; ,
\label{Int}
\end{equation}
where $G_H$ is a coupling constant which is related  through
bosonizations to physical quantities like $\alpha_H$ and $g$ appearing
 in (\ref{eq-8}) and (\ref{JH}), as well as $f_\pi$ and $m_\chi$.
 (See Appendix C).

Within HQEFT the heavy-light weak current in (\ref{Lcur}) will,
  below the renormalization scale $\mu = m_c$,
be modified in the following way \cite{neu}:
\begin{equation}
J_a^\lambda
=C_{\gamma}(\mu ) \bar{\chi}_b \xi^{\dagger}_{ba}\gamma^\lambda
L Q_v \, +
\, C_v(\mu ) \bar{\chi}_b \xi^{\dagger}_{ba}  v^\lambda L Q_v \; ,
\label{modcur}
\end{equation}
where the coefficients $C_{\gamma,v}$ are determined
by QCD renormalization for  $\mu < m_c$.
However, for $\mu \simeq \Lambda_\chi$,  $C_\gamma \simeq 1$
 and $C_v \simeq 0$. The bosonization of (\ref{modcur}) will lead to
(\ref{JH}) by using (\ref{Int}).

\section{Bosonization}

The Lagrangian \eqref{totlag} from the previous section can now
be used for bosonization, i.e. we
integrate out the quark fields. This can be done in the path
integral formalism, or as we do here,
by expanding in terms of  Feynman diagrams.
For instance, the lowest order (kinetic) chiral
 Lagrangian \eqref{eq-11} in the light
sector (involving $\pi, K,\eta$'s)
can be obtained
by coupling two axial fields to a quark loop using the Lagrangian
in Eq. \eqref{chqmR}:
\begin{equation}
 i {\cal{L}}^{(2)}_{str}(\pi, K,\eta) \,
= - N_c \, \int \frac{d^dp}{(2\pi)^d} \,
Tr \, \left[ \left( \gamma_\sigma \gamma_5 {\cal{A}}^\sigma \right)
 \, S(p) \,
\left( \gamma_\rho \gamma_5 {\cal{A}}^\rho \right)  \,  S(p) \right]
\sim
Tr \, \left[{\cal{A}}_\mu {\cal{A}}^\mu \right] \;,
\label{L2str}
\end{equation}
where $S(p) = ( \gamma \cdot p - m_\chi)^{-1}$, and the trace is
both in flavor and Dirac spaces.
This is  the standard form of the lowest order chiral Lagrangian
\eqref{eq-11}, which can  easily be seen by using the relations
\begin{equation}
{\cal{A}}_\mu \; = \; - \frac{1}{2 i} \xi \,
 (\partial_\mu \Sigma^\dagger) \, \xi
\; = \; \frac{1}{2 i} \xi^\dagger \, (\partial_\mu \Sigma) \,
\xi^\dagger \; .
\label{ASigma}
\end{equation}
Similarly one obtains
 the lowest order ${\cal O}(p)$
strong chiral Lagrangian \eqref{eq-8} in the heavy sector .

\begin{figure}[bt]
\begin{center}
  \epsfig{file=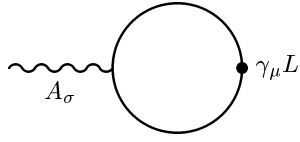,width=4cm}
\caption{Feynman diagram for bosonization of left-handed current
to order ${\cal O}(p)$}\label{fig:va}
\end{center}
\end{figure}

Let us now consider the bosonization of the pure light weak current.
The  lowest order term ${\cal O}(p)$ is obtained when the vertex
$\Lambda^X$
from
(\ref{leftcur}) and axial vertex ($ \sim {\cal A}_\mu$) from
(\ref{chqmR})
are combined with quark loops (see Fig. \ref{fig:va}):
\begin{equation}
 j^X_\mu({\cal{A}}) \, = - \, i N_c \int \frac{d^dp}{(2\pi)^d} \,
Tr \, \left[ \left(\gamma_\mu L \, \Lambda^X \right)  S(p) \,
\left( \gamma_\sigma \gamma_5 {\cal{A}}^\sigma \right)  \,  S(p) \,
\right]
  \;  \sim \; Tr \, \left[\Lambda^X \, {\cal{A}}_\mu  \right] \;.
\label{jXA}
\end{equation}
This coincides with (\ref{jX}) when (\ref{ASigma})
is used.

 To obtain a nonzero nonfactorizable contribution to
$D^0 \rightarrow K^0 \bar{K^0}$ at tree level,
 we have to consider the color current
$j^{X,a}_\mu$ to ${\cal O}(p^3)$, involving insertions of the
 ``mass fields'' $\widetilde{\cal{M}}_q$ in (\ref{cmass}).
 From Fig.~\ref{fig:colcur},
 one obtains the contribution:
\begin{figure}[bt]
\begin{center}
 \epsfig{file=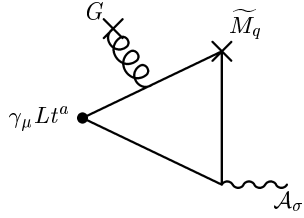,width=4cm}
\caption{Diagram for bosonozation of the color current to ${\cal O}(p^3)$}
\label{fig:colcur}
\end{center}
\end{figure}
\begin{equation}
 j^{X,a}_\mu(G^b,{\cal{A}},\widetilde{{\cal{M}}_q}|\;{\mbox{Fig. 7}}) \,
= i \, \int \frac{d^dp}{(2\pi)^d} \,
Tr \, \left[ \left(\gamma_\mu L \, \Lambda^X \right) \, S(p) \,
\left(\gamma_\sigma \gamma_5 {\cal{A}}^\sigma \right)  \, S(p) \,
 \widetilde{{\cal{M}}_q} \, S_1(p,G^b) \right]
  \,  ,
\label{jXG?}
\end{equation}
where
\begin{equation}
S_1(p,G^b) = - \frac{g_s}{4} G^b_{\alpha \beta} t^b
\left[ \sigma^{\alpha \beta} (\gamma \cdot p + m_\chi) \, + \,
(\gamma \cdot p + m_\chi) \sigma^{\alpha \beta}\right]
(p^2-m_\chi^2)^{-2}
\label{S1G}
\end{equation}
is the light quark propagator in a gluonic background (to first order
in the gluon field) and $g_s$ is the strong coupling constant. Moreover,
$a,b$
are color octet indices.
Summing all six diagrams with permutated vertices compared to the one in
Fig. \ref{fig:colcur} we obtain in total:
\begin{equation}
 j^{X,a}_\mu(G^b,{\cal{A}},\widetilde{{\cal{M}}_q}) \,
=  \,  \frac{g_s}{12 m_\chi} \frac{1}{16 \pi^2} \, G^{a, \kappa \lambda}
\left[ i \varepsilon_{\mu \rho \kappa \lambda} \;
T_{\varepsilon}^{X,\rho}
\, + \, \left(\eta_{\mu \kappa} \eta_{\rho \lambda} \; - \;
\eta_{\mu \lambda} \eta_{\rho \kappa}  \right) T_g^{X,\rho} \right] \; ,
\label{jXG}
\end{equation}
where (We have used the analytical
computer program FORM \cite{FORM})
\begin{equation}
T_{\varepsilon}^{X,\rho} \, = \, 4 \, S^K_\rho \, - 3 (S^L_\rho +
S^R_\rho)
\; , \;\;
T_g^{X,\rho} \, =  S^L_\rho \, - \,  S^R_\rho \; .
\label{Teg}
\end{equation}
The $S's$ are chiral Lagrangian terms:
\begin{equation}
\begin{split}
S^L_\rho \, & \equiv \, Tr \left[ \Lambda^X  {\cal{A}}_\rho \,
 \widetilde{{\cal{M}}_q}^L \right]
  \,  = \, \frac{1}{2i} Tr \left[ \lambda^X (D_{\rho} \Sigma)
\, {\cal{M}}_{q}^{\dagger}  \right]  \; ,  \\
S^R_\rho \, & \equiv \, Tr \left[ \Lambda^X \, \widetilde{{\cal{M}}_q}^R
 \, {\cal{A}}^\rho    \, \right]
  \,  = \, \frac{-1}{2i} Tr \left[ \lambda^X \, {\cal{M}}_{q} \,
(D_{\rho} \Sigma^\dagger ) \,  \right] \; ,  \\
S^K_\rho \, & \equiv \, \frac{1}{2}Tr \left[ \Lambda^X
\left(  {\cal{A}}^\rho \,  \widetilde{{\cal{M}}_q}^R
\, + \,  \widetilde{{\cal{M}}_q}^L  \, {\cal{A}}^\rho \right)   \,
\right]
  \\
 &= \, \frac{1}{4i} Tr \left[ \lambda^X \left( (D_{\rho} \Sigma)
\, \Sigma^\dagger {\cal{M}}_{q} \Sigma^\dagger \, - \,
 \Sigma {\cal{M}}_{q}^\dagger  \Sigma  (D_{\rho} \Sigma^\dagger)
\right)  \right] \; .
\label{SLRK}
\end{split}
\end{equation}

\vspace{0.2cm}

Within the heavy-light sector,
the weak current can be bosonized to lowest order (${\cal O}(p^0)$)
 by calculating the
Feynman diagram shown in Fig. \ref{fig:heavylight}, left.
The obtained result is Eq. (\ref{JH}) with $\alpha_H$ related to $G_H$
(see Appendix C).

\begin{figure}[bt]
\begin{center}
   \epsfig{file=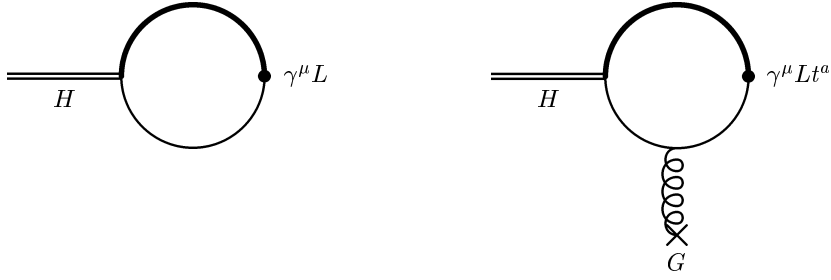,width=11cm}
\caption{Diagrams representing bosonization of heavy-light weak current.
The boldface line represents the heavy quark, the solid line the light
quark.}
\label{fig:heavylight}
\end{center}
\end{figure}

The
bosonization of the color current given by (\ref{modcur})
with an extra  color matrix $t^a$ inserted and with an extra gluon
emitted
is given by the following loop integral (Fig. \ref{fig:heavylight},
right):
\begin{equation}
J^\sigma(H_v \, G^a)^f \; = -
\int\,\frac{d^dk}{(2\pi)^d} Tr \left[ (-iG_H \, H_v \xi^{\dagger})^f
 (i S_1(k,G^b))  (\gamma^\sigma L t^a) (i \Delta_v(k)) \right] \; ,
\label{curGA}
\end{equation}
where  $\Delta_v(k) = P_+/k\cdot v$ is the heavy quark propagator.
 Notice that emission of a gluon from the heavy quark is suppressed by
$1/m_Q$
and omitted.
The result can be written
\begin{equation}
J^\sigma (H_v \, G^a)^f \; = \;  G_H \, g_s \, G^a_{\alpha \beta}
Tr \left[ \gamma_\sigma L \; (H_v \xi^\dagger)^f
\left( I_{G1} \, \sigma^{\alpha \beta}
 \, -i \, I_{G2}
(\gamma^{\alpha} v^{\beta} -  \gamma^{\beta} v^{\alpha} )\right) \right]
\; ,
\label{curGD}
\end{equation}
where $I_{G1}$ and $I_{G2}$ are loop integrals given
 in Appendix C.
Keeping only the pseudoscalar field $P_5$ representing $D^0$, we find
\begin{equation}
J^{Y,a}_\mu(P_5,G^b) \; = \; \frac{g_s \, G_H}{16 \pi^2}
 (P_5 \xi^\dagger)^Y  \,   G^{a,\alpha \beta}
\, \left[i B_\varepsilon \, \varepsilon_{\mu \sigma \alpha \beta}
v^\sigma
\; + \, B_g \left( \eta_{\mu \alpha} v_\beta \, - \, \eta_{\mu \beta}
v_\alpha
\right) \right] \; ,
\label{JYa}
\end{equation}
where $B_{\varepsilon, g}$ are obtained from loop integrals in
(\ref{curGD}).
Then we find the nonfactorizable (gluon condensate) contribution:
\begin{eqnarray}
 {\cal{L}}_{eff}(D^0 \mbox{decay})_{\langle G^2\rangle }
\; = \; 2 \widetilde{G} \,  c_A
\left( \frac{g_s \, G_H}{16 \pi^2} \right)
\left( \frac{g_s}{12 m_\chi} \frac{1}{16 \pi^2} \right) \langle
G^2\rangle  \nonumber \\
\times \, v_\rho
\, \left[ B_\varepsilon \, T^{X,\rho}_\varepsilon
\; + \, B_g \, T^{X,\rho}_g  \right] \,
  (P_5 \xi^\dagger)^Y  \; ,
\label{effLag}
\end{eqnarray}
where $\langle G^2\rangle $ is the gluon condensate, obtained by the
prescription
\begin{equation}
G_{\mu \nu}^a G_{\alpha \beta}^a \; \rightarrow \; \frac{1}{12} \,
 (\eta_{\mu \alpha} \, \eta_{\nu \beta} \, -
\, \eta_{\mu \beta} \,  \eta_{\nu \alpha}) \, \langle G^2\rangle  \; .
\end{equation}
In order to make predictions, we have to relate $G_H$ in (\ref{effLag})
and the various loop integrals to physical quantities like $m_\chi$,
$f_\pi$
and $\alpha_H \simeq f_D \sqrt{m_D}$.

It should be noted that there are apriori other terms than the
 one in (\ref{effLag}). There is one possible term where the field
$\widetilde{{\cal M}_q}$ occurring in Fig.~7 may instead be attached to
the light quark line in diagram in Fig.~8 (right). However, this term
will
not give contributions to $D^0 \rightarrow K^0 \bar{K^0}$. Moreover,
there
is apriori a term where the field ${\cal A_\sigma}$ attached
 in Fig.~7 is instead
attached to the light quark line in Fig.~8 (right). This term
is identically zero.

In the language of chiral perturbation theory, the term (\ref{effLag})
can be interpreted as a counterterm. To be more specific, the
(divergent part of the) counterterm
has the Lorentz and flavor structure of the second line of
(\ref{effLag})
and is multiplied with a (divergent) coefficient adjusted to cancel the
loop divergences obtained in Sect. 3.

\section{Results}

In our numerical calculation  we use the values of $\alpha_H$, $g$ and
$f$ obtained within the same framework in  \cite{stewart,BG,itchpt,
Grinstein-92,
Grinstein-94}.
The coupling $g$ is extracted from existing
experimental data  on $D^*\to D\pi$ and $D^*\to
D\gamma$ decays. This analysis \cite{stewart} includes chiral
corrections at
one
loop order and  yields
$g=0.27\genfrac {}{}{0pt}{}{+0.04+0.05}{-0.02-0.02}$, leaving
the sign undetermined.
The one loop chiral corrections  reduce the  pion decay constant from
$f_{\pi}=0.132$ $ {\rm GeV}$ to $f=0.120$ $ {\rm GeV}$
\cite{stewart}. In order
to obtain the $\alpha_H$ coupling, we use present experimental data on
$D_s$ leptonic decays.  Namely, at the tree level there is a relation
$f_{D}=f_{D_{s}}=\alpha_H/\sqrt{m_D}$. This relation receives $10 -
20\%$
chiral  corrections \cite{itchpt},
\cite{Grinstein-92}. From the experimental branching ratio
$D_s\to\mu \nu_\mu$ and the  $D_s$ decay width \cite{PDG-00} one gets
$f_{D_s}=0.23\pm 0.05$ $ {\rm GeV}$ and taking into account chiral
loop contributions,  we  find  $\alpha_H=0.23\pm0.04$ $ {\rm
GeV}^{3/2}$ (see
Appendix \ref{app-B}). Note  that  in \cite{itchpt} the
$\bar{\Delta}=0$ has been  used,  while we use
the strict $\overline{MS}$ prescription
 $\bar{\Delta}=1$ as in \cite{stewart}.
We put everywhere  $\mu = 1$ $\rm{GeV} \simeq \Lambda_\chi$.

For the Wilson coefficients $c_{A,B}$ of (\ref{Lquark})
 we use $c_A=1.10\pm 0.05$ and $c_B=-0.06\pm 0.12$
\cite{buras}, calculated   at the scale $\mu=1$ $
\rm{GeV}$ with the number of colors $N_c=3$.
Within the framework of ``new'' or ``generalized'' factorization, where
nonfactorizable effects are taken into account in a phenomenological
way,
one uses the ``effective values'' $c^{eff}_A=1.26$ and
$c^{eff}_B=-0.47$. However, in this paper we calculate nonfactorizable
effects in terms of chiral loops and gluon condensates, and therefore
we use the values of \cite{buras}.
 Due to the
 suppression of  $c_B$ in comparison with $c_A$, we do not include
 terms proportional to
$c_B$. \footnote{Even if the ``new factorization'' values had been used,
the $c_B$ part of weak interaction would be  suppressed  by $1/3$
compared to the $c_A$ one.}
We present our numerical results for the nonzero amplitudes in Table
 \ref{tab-1}.

\begin{table} [h]
\begin{center}
\begin{tabular}{|l|c|} \hline
$-$&${\cal M}_i[\times 10 ^{-7}{\rm  \;GeV}]$\\ \hline\hline
${\cal M}_1$ & $-0.42$ \\ \hline
${\cal M}_2$ & $-0.31$ \\ \hline
${\cal M}_3$ &$ -0.62$ \\ \hline
${\cal M}_4$ & $0.28 -2.44 i$ \\ \hline
${\cal M}_5$ &$-0.81$ \\ \hline
${\cal M}_6$ &$-0.61$ \\ \hline
${\cal M}_{7} $&$-0.99$ \\ \hline
${\cal M}_{8}$ &$ 0.92$ \\ \hline\hline
$\sum_i {\cal M}_i$& $-2.56 -2.44 i$\\ \hline

\end{tabular}
\caption{\footnotesize{Table of the one chiral loop  amplitudes (see
Fig.
\ref{fig-5}), where ${\cal M}=\sum_n {\cal M}$ is defined in
\eqref{eq-12}.
In the last line the sum of all amplitudes is presented. It can be
compared
with the experimental result 
$|{\cal M}_{\rm Exp}|=3.80\times 10^{-7}{\rm \;GeV}$.}}
\label{tab-1}
\end{center}
\end{table}

The imaginary
part of the amplitude comes from the $F_4$ graph, when  the  $\pi$'s
or the  $K$'s  in the loops are on-shell. All other
 graphs contribute only to the
real part of the amplitude.
The imaginary part of the amplitude is scale and scheme independent
within chiral perturbation theory.
This amplitude is also obtained from unitarity, and is valid beyond
the chiral loop expansion.
We also mention that the rescattering contribution, considered in
\cite{XYP,BLMPS}
is  the same contribution as the one we calculate  from graphs on
Fig. \ref{fig-1}.
% if $D^0 \to K^+ K^-(\pi^+ \pi^-)$.

In order to cancel divergences one has to construct counterterms.
In our case, this is described at the end of section 5.
Generally, one can do that by using the symmetry arguments,
as it has been done in \cite{BG,stewart} for the semileptonic decays
of heavy mesons and $D^*$ decays. In  the case of $D^*$ \cite{stewart}
it was
estimated that the contribution of counterterms is not substantial.

To obtain the $D^0 \rightarrow \bar{K^0} K^0$ amplitude due to
 gluon condensate
 we have to know the coupling $G_H$. In addition, we have to find
the tensors $S$ in (\ref{Teg}, \ref{SLRK}) (and thereby the $T$'s) for
  $K^0 \bar{K^0}$ in the final state. We find to lowest order
for the parts of
$T^{X,\rho}_{g,\varepsilon}$:
\begin{equation}
S_\mu^L \, = \, - \, S_\mu^R \, =
\; -  \frac{1}{f^2} (m_s -m_d) \, (p + \bar{p})_\mu   \; ; \quad
S^K_\mu \, = \, \frac{2}{f^2} (m_s + m_d) \, (p -  \bar{p})_\mu \, \; ,
\label{Vkkbar}
\end{equation}
where $p$ and $\bar{p}$ are the momenta of $K^0$ and $\bar{K^0}$
respectively. From (\ref{JYa}) we  see,  that the momenta will be
contracted with
$v^\mu = p_D^\mu /M_D$; where $p_D = p +\bar{p}$.
It is important that $S_\mu^L$ and  $S_\mu^R$ have a different momentum
structure than $\langle K^0 \bar{K^0}|V_\mu|0\rangle $ in Eq.
(\ref{eq-9}), and
they  will
give a nonfactorizable contribution to $D^0 \rightarrow K^0 \bar{K^0}$
 proportional to $\langle G^2\rangle $, while $S_\mu^K$ does not.
 Note that $T^{X,\rho}_{\varepsilon}$ of Eqs.
 (\ref{jXG},\ref{Teg},\ref{effLag}) do not contribute. We find the
gluon condensate contribution:
\begin{equation}
{\cal{M}}(D^0 \rightarrow K^0 \bar{K^0})_{\langle G^2\rangle }
\; = \;  c_A \, (\widetilde{G} m_D^2) \, \frac{(m_s - m_d)}{m_\chi} \;
\frac{\beta \, \delta_G}{6 N_c} \, B_g \, f_D
\label{amp}
\end{equation}
where:
\begin{equation}
\delta_G \equiv  N_c \frac{\langle \alpha_s \, G^2/\pi\rangle }{8 \pi^2
f^4} \; ,
\quad
G_H \equiv \beta \, \frac{\alpha_H}{f^2} \; , \quad
B_g = 16 i \pi^2 (I_{G1}- I_{G2}) = \frac{\pi}{4} \;.
\label{DivRel}
\end{equation}
 When we  take into account
 the various relations between the loop integrals ($I$'s) and $G_H$,
 we find that $\beta \simeq 1/4$.
Using the values \cite{BEF}
 $\langle \frac{\alpha_s}{\pi} G^2\rangle  \simeq (333 {\mbox \;{\rm
MeV}})^4$, $m_\chi=200$
\;MeV,
and  $m_s \simeq$  150 \;MeV, we obtain the numerical value:
\begin{equation}
{\cal{M}}(D^0 \rightarrow K^0 \bar{K^0})_{\langle G^2\rangle }
\; \simeq \;  0.87 \times 10^{-7}  \text{\;GeV} \; ;
\label{ampNum}
\end{equation}
which is also of the same order of magnitude as the experimental value. 

Adding both the  chiral loops and the gluon
condensate contributions,  we obtain the total amplitude to ${\cal
O}(p^3)$
\begin{equation}
{\cal M}_{\text{Th}}=(-1.7-2.4 \, i)\times 10^{-7}{\text{\;GeV}} \; ,
\end{equation}
or in terms of branching ratio
\begin{equation}
B(D^0\to K^0\bar{K^0})_{\text{Th}}=(4.3\pm 1.4)\times 10^{-4} \;,
\end{equation}
where the estimated  uncertainties reflect the  uncertainties
 in the input parameters, especially $\alpha_H$. 

Around the charm mesons mass region there are many resonances.
One might think that their contribution will appear in this decay mode,
 either as scalar resonance exchange like in \cite{DDL}
or as $K^*$  exchanges \cite{DDL,Z,BLMPS}.
Within our framework they would appear as the next
order contribution (${\cal O}(p^5)$) in the chiral expansion .
This is,  however,  beyond the present scope of our investigations.
It is interesting to point out that the effects we calculate,
both from chiral loops and from the gluon condensate,
 are results of the $SU(3)$ flavor
symmetry breaking. In the limit of exact symmetry both contributions
will
disappear.

We can summarize that we  indicate the leading ${\cal O}(p^3)$
 nonfactorizable
contributions to $D^0 \to K^0 \bar K^0$. The chiral loops and gluon
condensates give the contributions of the same order of magnitude
as the amplitude extracted from the experimental result.

\appendix

\section{List of integrals from chiral loops}

\label{app-A}
Here we list the dimensionally regularized integrals needed for
evaluation of $\chi$PT and HQEFT one-loop graphs shown on Fig.
\ref{fig-5}:
\begin{align}
i\mu^\epsilon\int \frac{d^{4-\epsilon}q}{(2\pi)^{4-\epsilon}
}\frac{1}{q^2-m^2}&=\frac{1}{16 \pi^2}I_1(m),\\
i\mu^\epsilon\int \frac{d^{4-\epsilon}q}{(2\pi)^{4-\epsilon}}
\frac{1}{(q^2-m^2)(q\cdot v-\Delta)}&=\frac{1}{16
\pi^2}\frac{1}{\Delta}I_2(m,\Delta),
\end{align}
with
\begin{align}
I_1(m)&=m^2 \ln\Bigl(\frac{m^2}{\mu^2}\Bigr)-m^2\bar{\Delta} ,\\
I_2(m,\Delta)&=-2\Delta^2 \ln\Bigl(\frac{m^2}{\mu^2}\Bigr)-4 \Delta^2
F\Bigl(\frac{m}{\Delta}\Bigr) +2 \Delta^2(1+\bar{\Delta}),
\end{align}
where $\bar{\Delta}=\frac{2}{\epsilon}-\gamma +\ln(4\pi)+1$ (in
calculation $\bar{\Delta}=1$), while $F(x)$ is the function calculated
by Stewart in \cite{stewart}, valid for negative and positive values of
the argument
\begin{equation}
F\left(\frac{1}{x}\right)= \left\{
\begin{aligned}
-\frac{\sqrt{1-x^2}}{x}&\left[\frac{\pi}{2}-\tan^{-1}\left(\frac{x}{\sqrt{1-x^2}}\right)\right]\qquad
&|x|\le 1\\
\frac{\sqrt{x^2-1}}{x}&\ln \left(x+\sqrt{x^2-1}\right)\qquad &|x|\ge 1
\; .
\end{aligned}\right.
\end{equation}
The other integrals needed are
\begin{equation}
i\mu^\epsilon\int \frac{d^{4-\epsilon}q}{(2\pi)^{4-\epsilon}}
\frac{q^\mu}{(q^2-m^2)(q\cdot v-\Delta)}=\frac{v^\mu}{16
\pi^2}[I_2(m,\Delta)+I_1(m)],
\end{equation}
\begin{equation}
i\mu^\epsilon\int \frac{d^{4-\epsilon}q}{(2\pi)^{4-\epsilon}}
\frac{q^\mu q^\nu}{(q^2-m^2)(q\cdot v-\Delta)}=\frac{1}{16
\pi^2}\Delta\left[ J_1(m,\Delta)\eta^{\mu \nu}+J_2(m,\Delta)v^\mu
v^\nu\right],
\end{equation}
with
\begin{subequations}\label{eq-1}
\begin{align}
\begin{split}
J_1(m,\Delta)=(-m^2&+\frac{2}{3}\Delta^2)\ln\left(\frac{m^2}{\mu^2}\right)+\frac{4}{3}(\Delta^2-m^2)F\left(\frac{m}{\Delta}\right)\\
&\qquad\qquad -
\frac{2}{3}\Delta^2(1+\bar{\Delta})+\frac{1}{3}m^2(2+3\bar{\Delta})+\frac{2}{3}m^2-\frac{4}{9}\Delta^2
\; ,
\end{split}
\\
\begin{split}
J_2(m,\Delta)=(2
m^2&-\frac{8}{3}\Delta^2)\ln\left(\frac{m^2}{\mu^2}\right)-\frac{4}{3}(4
\Delta^2-m^2)F\left(\frac{m}{\Delta}\right)\\
&\qquad\qquad+\frac{8}{3}\Delta^2(1+\bar{\Delta})-\frac{2}{3}m^2(1+3\bar{\Delta})-\frac{2}{3}m^2+\frac{4}{9}\Delta^2
\; ,
\end{split}
\end{align}
\end{subequations}
The functions $J_1(m,\Delta),J_2(m,\Delta)$ differ from the ones in
Boyd - Grinstein list of integrals \cite{BG} by the last two terms in
\eqref{eq-1} that are of the order  of ${\cal O}(m^2, \Delta^2)$. These
additional finite terms originate from the fact that $\eta^{\mu \nu}$ is
$4-\epsilon$ dimensional metric tensor.

The chiral loop integrals needed are
\begin{align}
i \mu^\epsilon\int
\frac{d^{4-\epsilon}q}{(2\pi)^{4-\epsilon}}\frac{1}{((q+k)^2-m^2)(q^2-m^2)}&=\frac{1}{16
\pi^2} N_0(m,k^2),\\
i \mu^\epsilon\int
\frac{d^{4-\epsilon}q}{(2\pi)^{4-\epsilon}}\frac{q^\mu}{((q+k)^2-m^2)(q^2-m^2)}&=\frac{k^\mu}{16
\pi^2}N_1(m,k^2)=-\frac{1}{2}\frac{k^\mu}{16 \pi^2}N_0(m,k^2),\\
i \mu^\epsilon\int
\frac{d^{4-\epsilon}q}{(2\pi)^{4-\epsilon}}\frac{q^\mu
q^\nu}{((q+k)^2-m^2)(q^2-m^2)}&=-\frac{k^\mu k^\nu}{16 \pi^2}
N_2(m,k^2)-\frac{\eta^{\mu\nu}}{16 \pi^2}N_3(m,k^2),
\end{align}
where
\begin{align}
N_0(m,k^2)&=-\bar{\Delta}+1-H\Bigl(\frac{k^2}{m^2}\Bigr)+\ln\Bigl|\frac{m^2}{\mu^2}\Bigr|-i
\pi \Theta\Bigl(-\frac{m^2}{\mu^2}\Bigr)\sign(\mu^2)\; ,\\
\begin{split}
N_2(m,k^2)&=\frac{1}{3}\left[\bar{\Delta}+\frac{7}{6}-2\frac{m^2}{k^2}+2\biggl(\frac{m^2}{k^2}-1\biggr)\biggl(1-\frac{1}{2}H\Bigl(\frac{k^2}{m^2}\Bigr)\biggr)\right.\\
&\left.\qquad\qquad\qquad -\ln\Bigl(\frac{m^2}{\mu^2}\Bigr)+i \pi
\Theta\Bigl( -\frac{m^2}{\mu^2}\Bigr)\sign(\mu^2)\right]\; ,
\end{split}\\
\begin{split}
N_3(m,k^2)&=\frac{1}{2}\Bigl(m^2-\frac{k^2}{6}\Bigr)\bar{\Delta}-\frac{1}{2}\Bigl\{\frac{1}{3}\bigl(8
m^2+k^2)\biggl[
1-\frac{1}{2}H\Bigl(\frac{k^2}{m^2}\Bigr)\biggr]-\frac{8}{3}m^2\Bigr.\\
&\Bigl.\qquad\qquad\qquad-\frac{5}{18}k^2+\Bigl(
m^2+\frac{k^2}{6}\Bigr)\Bigl(\ln\Bigl|\frac{m^2}{\mu^2}\Bigr|-i\pi
\Theta\Bigl(-\frac{m^2}{\mu^2}\Bigr)\sign(\mu^2)\Bigr)\Bigr\} \; ,
\end{split}
\end{align}
and
\begin{equation}
H(a)=\left\{
\begin{aligned}
2&\left(1-\sqrt{4/a-1}\arctan\left(\frac{1}{\sqrt{4/a-1}}\right)\right)\qquad
& 0<a<4\\
2 &
\left(1-\frac{1}{2}\sqrt{1-4/a}\left[\ln\left|\frac{\sqrt{1-4/a}+1}{\sqrt{1-4/a}-1}\right|-i\pi\Theta(a-4)\right]\right)\qquad
& \text{ otherwise}
\end{aligned}
\right.
\end{equation}
while $m^2$ is assumed to be positive.

\section{D meson decay constant}\label{app-B}
Here we list results for one-loop chiral corrections to $D$ meson decay
constants and use them to obtain coupling $\alpha_H$ from experimental
data.  The one-loop chiral corrections have been calculated  in
\cite{itchpt}, \cite{BG} using $\bar{\Delta}=0$, while the leading logs
have been obtained already in \cite{Grinstein-92}, \cite{Goity-92}
\begin{subequations}
\begin{align}
\begin{split}
f_D=\frac{\alpha_H}{\sqrt{m_D}}\Big[& 1+ \frac{3
g^2}{32\pi^2f^2}\Big(\frac{3}{2}
C(\Delta_{D^*D},m_\pi)+C(\Delta_{D_s^*D},m_K)+\frac{1}{6}C(\Delta_{D^*D},m_\eta)\Big)\\
&-\frac{1}{32\pi^2f^2}\Big(\frac{3}{2}I_1(m_\pi)+I_1(m_K)+\frac{1}{6}I_1(m_\eta)\Big)\Big],
\end{split}
\\
\begin{split}
f_{D_s}=\frac{\alpha_H}{\sqrt{m_D}}\Big[& 1+ \frac{3
g^2}{32\pi^2f^2}\Big(2
C(\Delta_{D^*D_s},m_K)+\frac{2}{3}C(\Delta_{D_s^*D_s},m_\eta)\Big)\\
&-\frac{1}{32\pi^2f^2}\Big(2I_1(m_K)+\frac{2}{3}I_1(m_\eta)\Big)\Big],
\end{split}
\end{align}
\end{subequations}
where $C(\Delta,m)=J_1(m,\Delta)+\Delta \frac{\partial}{\partial
\Delta}J_1(m,\Delta)$, while $J_1(m,\Delta)$ and $I_1(m)$ can be found
in appendix \ref{app-A}. Using $f=120{\rm MeV}$, $\mu=1 {\rm GeV}$ and
$\bar{\Delta}=1$ one gets the numerical values
\begin{subequations}
\begin{align}
f_{D}=&\frac{\alpha_H}{\sqrt{m_D}}(1+0.18-0.37g^2)\label{eq-7} \; ,\\
f_{D_s}=&\frac{\alpha_H}{\sqrt{m_D}}(1+0.35 +0.38g^2)\label{eq-6} \; .
\end{align}
\end{subequations}

To obtain the $\alpha_H$ coupling we use experimental data on decays of
$D$ mesons into leptons. From the experimental value for branching ratio
$B(D_s\to\mu \nu_\mu)=(4.6\pm1.9) 10^{-3}$ and the  $D_s$ decay time
$\tau_{D_s}=(0.496 \genfrac{}{}{0pt}{}{+0.010}{-0.009})\cdot 10^{-12} s
$ one gets $f_{D_s}=0.23\pm 0.05 $ ${\rm GeV}$.
Using this value and
$g=0.27$ \cite{stewart} in \eqref{eq-6} we get
$\alpha_H=0.23\pm0.04$ ${\rm GeV}^{3/2}$.

Using $\alpha_H=0.23\pm0.04$ $ {\rm GeV}^{3/2}$ in  \eqref{eq-7} and
$g=0.27$
we can also calculate $f_{D}=0.194\pm 0.045 $ ${\rm GeV}$, where the
uncertainties
are due to  the uncertainties in $\alpha_H$. The average value for the
ratio
$f_{D_s}/f_{D}=1.19$ is in fair agreement with the recent lattice
results \cite{Becirevic-00}.

\section{Heavy-light quark loop integrals}\label{app:form}

 The integrals entering heavy quark loops like the ones in Fig.8
 are of the form:
\begin{equation}
R_{p,q} \equiv \int\,\frac{d^dk}{(2\pi)^d}
\frac{1}{(v.k)^p} \,  \frac{1}{(k^2-m^2)^q} \; .
\end{equation}
Performing a shift of momentum integration combined with Feynman
parameterization, we obtain
\begin{equation}
R_{p,q} \; = \; 2^p \, \frac{\Gamma(p+q)}{\Gamma(p) \Gamma(q)}
\; K(p+q,p-1) \; ,
\end{equation}
where
\begin{equation}
K(n,r) \equiv \int\limits_0^\infty\,d\lambda\int\,\frac{d^dl}{(2\pi)^d}
\frac{\lambda^r}{(l^2-m^2-\lambda^2)^n}\; .\label{eq-13}
\end{equation}

One should notice that to obtain the result in (\ref{eq-8}),
 we have to do the  identification
\begin{equation}
8 i N_c G_H^2 I_{HH} = 1 \; ,
\label{norm}
\end{equation}
where $I_{HH}$ is a logarithmically divergent loop integral
given below.
(There is also  a similar relation for $g$.)
One should notice that some authors use an extra factor $m_H$, the mass
of the
heavy meson, in front of the right hand side of (\ref{eq-8}).
 Choosing the normalization in (\ref{eq-8}), it means
 that a factor $\sqrt{m_H}$ is included in the heavy meson field $H_v$.
For the left handed current in (\ref{JH}) and (\ref{fD})
 we find that we have to identify :
\begin{equation}
\alpha_H = -4i N_c G_D \,  I_{HW}  \; ,
\label{fDnorm}
\end{equation}
where   $I_{HW}$ is a quadratically divergent loop integral.

 The regularization can be done in various ways (various cut-off
prescriptions or  by $\overline{MS}$)
 and each
regularization correspond to slightly different versions of this type
model
\cite{BEF,epb,barhi,effr,itCQM,AJ}. For instance, in the version we use,
when soft gluon emission is  included in (\ref{curGD}) above, gluon
condensate
contributions  should also
  be included in loop integrals $I_{HH}$ and $I_{HW}$, as it is
for $f_\pi$ in the light sector \cite{BEF,pider}. However, we will not
go into these details here.
 Still, as in the pure light quark case, one  obtains
 numbers in the right ball park by
parameterizing the quadratic divergent integral as $\Lambda_\chi^2$, and
the
logarithmically divergent integral as $log(\Lambda_\chi^2/m^2)$.
Anyway, using the  expression for $f_\pi$ obtained in the $\chi$QM,
we obtain from (\ref{norm}) to leading order
\begin{equation}
G_H \; \simeq \; \frac{2 \sqrt{m_\chi}}{f_\pi} \; .
\label{GHlog}
\end{equation}
It can be seen from Ward identities for the loop diagrams
for, say Fig.~8 (left) that the quadratic
divergence in $I_{HW}$ is related to the quark condensate of the light
quark, which is also quadratically divergent. Then, similar to
(\ref{GHlog}),
we obtain from (\ref{fDnorm}) to leading order
\begin{equation}
G_H \; \simeq \;  -2 \frac{m_\chi \, \alpha_H}{\langle \bar{q} q\rangle
}   \; ,
\label{GHquad}
\end{equation}
Combining (\ref{GHlog}) and (\ref{GHquad}) we obtain
\begin{equation}
\alpha_H \; \simeq \; - \, \frac{\langle \bar{q}q\rangle }{f_\pi \,
\sqrt{m_\chi}} \; ,
\label{alphaH}
\end{equation}
which for the values $m_\chi$= 200 MeV, $f_\pi$ = 131 MeV and
 $\langle \bar{q}q\rangle $ = (-240MeV$)^3$ gives the value for
$\alpha_H$ cited in
Appendix B. Furthermore, using (\ref{GHlog}) and (\ref{GHquad})
we obtain
\begin{equation}
\beta \; \simeq \; - 2 \, \frac{m_\chi \, f_\pi^2}{\langle \bar{q}
q\rangle } \; \simeq
\; 0.25 \; ,
\label{beta}
\end{equation}
to be used in (\ref{amp}) and (\ref{DivRel}) .

Within dimensional regularization,
the expressions for some vales of $n$ and $r$ are listed below:
\begin{align}
K(2,1)&=\frac{i}{2{(4\pi)}^{d/2}}\frac{\Gamma (1-d/2)}{(m^2)^{1-d/2}} \;
,\\
K(3,1)&=-\frac{i}{4{(4\pi)}^{d/2}}\frac{\Gamma
(2-d/2)}{(m^2)^{2-d/2}}\label{k4}\; , \\
K(3,2)&=-\frac{i}{16{(4\pi)}^{d/2-1/2}}\frac{\Gamma
(3/2-d/2)}{(m^2)^{3/2-d/2}}
\; . \label{k5}
\end{align}
where $m=m_\chi$. From the properties of the $\Gamma$ function it is
easy to
see that:
\begin{equation}
K(2,0) = - 4 K(3,2) \quad ; \qquad   m^2K(3,0) = (3-d)K(3,2)\; ,
\label{k7}
\end{equation}
Comparing with a cut-off regularization, we see that $K(2,1)$ is
quadratically
and $K(3,1)$ is logarithmically divergent. In a primitive cut-off
regularization $K(2,0)$ and $K(3,2)$ appear as linearly divergent
\cite{barhi},
while they here appear as finite!

Note also that some of the integrals \eqref{eq-13} can be obtained as
the
limits of integrals listed in appendix \ref{app-A} if one lets $\Delta
\to 0$.
Thus one has the relations
\begin{align}
K(2,0)&=-\frac{i}{32\pi^2}\lim_{\Delta\to
0}\frac{1}{\Delta}I_2(m,\Delta)\; ,\\
K(2,1)&=\frac{i}{32 \pi^2}\lim_{\Delta\to
0}\Big[I_2(m,\Delta)+I_1(m)\Big]=\frac{i}{32 \pi^2}I_1(m)\; ,\\
K(2,2)&=-\frac{i}{32\pi^2}\lim_{\Delta\to 0}\Delta J_2(m,\Delta)\; .
\end{align}

The loop integral's (the $I$'s)  are defined as:
\begin{align}
I_{HH}& \equiv \quad mK(3,1)+K(3,2)\; ,\label{kappa1}\\
I_{HW}& \equiv \quad K(2,1)+mK(2,0)\; ,\\
I_{G1}& \equiv \quad K(3,1) + m K(3,0)\; ,\\
I_{G2}& \equiv \quad K(3,1)\; .\\
\label{Irel}
\end{align}

\end{fmffile}
\end{document}